\documentclass[twocolumn,showpacs,preprintnumbers,amsmath,amssymb]{revtex4}

\usepackage{graphicx}
\usepackage{bm}% bold math

\begin{document}

\title{ Block model for the XY-type Landau-Ginzburg-Wilson Hamiltonian with an inhomogeneous
temperature}

\author{X.T. Wu}

\affiliation{Department of Physics, Beijing Normal University,
Beijing, 100875, China}

\date{\today}% It is always \today, today,

\begin{abstract}
The phase fluctuation near the saddle point solution  of the
XY-type Landau-Ginzburg-Wilson Hamiltonian with random temperature
is studied. For the modes with lowest eigenvalue, the systems is
self-organized into blocks, which are coupled as a XY model with
random bond. The couplings obtained in this way agree with those
by domain wall method.
\end{abstract}

\pacs{05.70.-a,05.70.Fh+q, 64.60.-i, 64.60.Bd}

%Key words: Excited state solution, phase transition, disordered systems, Landau-Ginzburg Hamiltonian.

\maketitle

\section{Introduction}

In recent years more and more experiments show locally ordered
regions (LOR). Localized Bose-Einstein condensation above the
global superfluid transition temperature is revealed in superfluid
transition of $^4He$ in silica gel\cite{plan2,shirahama}. It is
well-known that for some granular superconductors, on the
insulating side of the superconductor insulator transition, each
grain is separately and independently superconducting while a
transport measurement shows the film to be insulating
\cite{merchant}. In recent experiment on amorphous NbN films,
pseudo-gapped state due to the locally superconducting islands is
discovered \cite{mondal}. The nucleation of pairing gaps in
nanoscale regions above $T_c$ is also found in the high-$T_c$
supercondcutor $Bi_2Sr_2CaCu_2O_{8+\delta}$ \cite{gomes}. In
addition, a local metallic state is observed in globally
insulating $La_{1.24}Sr_{1.76}Mn_2O_7$ well above the
metal-insulator transition \cite{sun}. The existence of
Ferromagnetic region in the paraphase of $La_{1-x}Ba_xMnO_3$ is
discovered \cite{eremina}.

To understand the relation between LOR and the phase transition in
a general way, the saddle point equation of Landau-Ginzburg
Hamiltonian with random temperature is solved recently
\cite{wu1,wu2}. LOR is explicitly shown in these solutions.
Moreover it is found that there exist many excited solutions,
which minimize the Hamiltonian locally in the configuration space.
These solutions can be described by the block model (in the
following we call it B model), in which the system is
self-organized into blocks. These blocks behave like superspins
and are coupled with their neighbors. In reference \cite{wu3}, a
general method to calculate the couplings between adjoined blocks
is proposed. This method is based on the free energy increasing of
domain wall (DW) between the adjoined blocks. So we call this
method DW method.

However for the systems with continuous order parameter, such as
superfluid, superconductor, the continuous phase fluctuation about
the saddle point solution should be taken into account. In DW
method this kind of fluctuation is absent. In this paper we study
the XY-type landau-Ginzburg-Wilson (LGW) Hamiltonian with random
temperature. We propose a Gaussian approximation to study the
continuous phase fluctuation near the saddle point solutions. We
will show that

(1). The system is still be organized into blocks given by the DW
method.

(2). For the modes with lower eigenvalues, the couplings between
the blocks are XY-type like a Josephson junction array and the
couplings are approximately equal to those given by the DW method.
The blocks are coupled like XY-type spins.

Our paper is arranged as follows. In section II, the model of
XY-type landau-Ginzburg-Wilson (LGW) Hamiltonian with random
temperature is given. In section III, one dimensional case is
discussed. In section IV, two dimensional case is discussed.
Section V is a summary.

\section{The model}

We consider the XY-type landau-Ginzburg-Wilson (LGW) Hamiltonian
with random temperature
\begin{equation}
H=\int d{\textbf r} \{{1 \over 2} |\bigtriangledown \phi({\textbf
r})|^2
   +{1 \over 2}t({\textbf r})\phi^2({\textbf r})
   +{1 \over 4}\phi^4({\textbf r})\},
\label{eq:h0}
\end{equation}
where
\begin{equation}
\phi=(\phi_x,\phi_y),\hskip 0.5cm \phi^2=\phi_x^2+\phi_y^2, \hskip
0.5cm |\bigtriangledown \phi|^2=|\bigtriangledown
\phi_x|^2+|\bigtriangledown \phi_y|^2,
\end{equation}
and $t({\textbf r})=t+\tilde{t}({\textbf r})$, and
$t,\tilde{t}({\textbf r})$  are the average reduced temperature
and the random part caused by the disorder respectively.  The
parameters are scaled according to the references \cite{wu3,wu1}.

The saddle point equations are given by
\begin{equation}
-\bigtriangledown^2 \phi_x({\textbf r}) +[t({\textbf r})
   + \phi^2 ({\textbf r})]\phi_x({\textbf r})=0,
\label{eq:spe1}
\end{equation}
\begin{equation}
-\bigtriangledown^2 \phi_y({\textbf r}) +[t({\textbf r})
   + \phi^2 ({\textbf r})]\phi_y({\textbf r})=0.
\label{eq:spe1}
\end{equation}

Through this paper we assume the saddle point solutions are along
$\phi_x$ direction,
\begin{equation}
\phi_x=\phi_x^{(\nu)},\hskip 0.5cm \phi_y=0.
\end{equation}
where $\nu$ is used to label the excited states.

Substituting the saddle point solution into Eq. (\ref{eq:h0}), one
get the free energy \cite{dots1}
\begin{equation}
F_{\nu}=H(\{\phi_x^{(\nu)}\})=-\int d{\textbf r} {1 \over 4}
(\phi_x^{(\nu)})^4({\textbf r}), \label{eq:fe}
\end{equation}
for the $\nu$th solution.

If we assume the solution is along $\phi_x$ direction and ignore
the fluctuation in  $\phi_y$ direction, the problem is reduced to
be Ising-type. It has been shown that in that case the system is
self-organized into blocks and the blocks are coupled like
Ising-spins \cite{wu2}. The couplings between blocks can be
obtained by calculating the free energy increase due to the domain
wall \cite{wu3}. If $(\phi_x,\phi_y)$ is regarded as a complex
parameter, letting $\phi_y=0$ is a constraint that only the phase
of $0$ and $\pi$ is allowed. If we take the fluctuation of
$\phi_y$ is into account, the phase fluctuation becomes
continuous.

In order to write the Hamiltonian in terms of the amplitude and
phase of the order parameter, we introduce
\begin{equation}
\phi_x=\Phi \cos \theta, \hskip 1 cm  \phi_y=\Phi \sin \theta.
\end{equation}
Then LGW Hamiltonian becomes
\begin{equation}
H=\int d{\textbf r} \{{1 \over 2} [|\bigtriangledown
\Phi|^2+\Phi^2|\bigtriangledown \theta|^2]
   +{1 \over 2}t({\textbf r})\Phi^2
   +{1 \over 4}\Phi^4\}.
\label{eq:h1}
\end{equation}
In this form, we can see that the free energy increase induced by
the variation of phase is proportioned to the square of amplitude.

\section{One-dimensional case}

\subsection{The method of domain wall}

As an example, we first consider a system with size being $14$
consisting of 7 wells and 7 barriers, i.e. the temperature field
is given by
\begin{equation}
t(\xi)=\left \{ \begin{array}{cc}
            t_b; & 2i-2<\xi\leq 2i-1, \\
            t_w; & 2i-1<\xi\leq 2i.
           \end{array} \right.
\label{eq:lattice}
\end{equation}
where $i=1,2,...,7$. Here we let the spatial coordinate be $\xi$
to distinguish from the directions of order parameter.

Obviously this system is periodic with period of $2$. The ground
state solution $\phi_x^{(0)}$ for the temperature field with
$t_w=-30,t_b=10$ is shown in the Fig. (1a). The saddle point
equation is solved by finite-difference method with step $h=0.025$
\cite{koon}. The saddle point solution is approximately equal to
$\sqrt{-t_w}$ at the centers of wells and decays to very small in
the barriers \cite{wu1}. Using the DW method, we can show that
there are $7$ elementary blocks and can calculate the couplings
between the adjoined blocks. For example, letting the initial
value of $\phi_x$ be negative in the first well and positive at
other positions, we will get the excited solutions $\phi_x^{(1)}$,
which is also shown in the Fig. (1a). The spatial range of the
first block is given by $\phi_x^{(1)}(\xi)<0$. Explicitly it is
$0.5<\xi\le 2.5$.

Similarly, letting the initial value of $\phi_x$ be negative in
the second well and positive at other positions, we will get the
excited solutions $\phi_x^{(2)}$. The spatial range of the second
block is $2.5<\xi\le 4.5$. We do not show it in figure since its
shape is the same as $\phi_x^{(1)}$ and can be obtained by
shifting $\phi_x^{(1)}$ in $\xi$-axis by $2$. Therefore the
spatial ranges of the 7 elementary blocks are given by
$2(i-1)+0.5<\xi\le 2i+0.5$, where satisfy $\phi_x^{(i)}<0$, for
$i=1,2,\cdots,9$.

Here we give a general method to get the spatial range of the
elementary block. For the general disordered cases, the domain
wall of elementary block can obtained by the method of opening
windows \cite{wu3}. The spatial range of this block is that
surrounded by the domain wall.

Letting the initial value of $\phi_x$ be negative in both the
first and second well and positive at other positions, we will get
the excited solutions $\phi_x^{(12)}$. Substituting the solutions
$\phi_x^{(0)},\phi_x^{(1)},\phi_x^{(2)},\phi_x^{(12)}$ into Eq.
(\ref{eq:fe}), we get the free energies $F_0,F_1,F_2,F_{12}$, and
the free energy increases $f_1=F_1-F_0$, $f_2=F_2-F_0$ and
$f_{12}=F_{12}-F_0$. Then the couplings between blocks is given by
\cite{wu3}
\begin{equation}
K_{12}^{(D)}=(f_1+ f_2- f_{12})/2.
\end{equation}
Here we use ``(D)" to denote the DW method. In this way we get the
couplings between adjoined blocks. Then the free energy of $\nu$th
state is given by
\begin{equation}
f_{\nu}=-\sum_{i}K_{i,i+1}(\sigma_i \sigma_{i+1}-1)/2
\label{eq:ising}
\end{equation}
where $\sigma_i$ is the sign of the ith block in the $\nu$th
solution.

The couplings between adjoined blocks with $t_w=-30$ at different
$t_b$ are given in table 1.

\begin{figure}
 \begin{center}
    \resizebox{8cm}{8cm}{\includegraphics{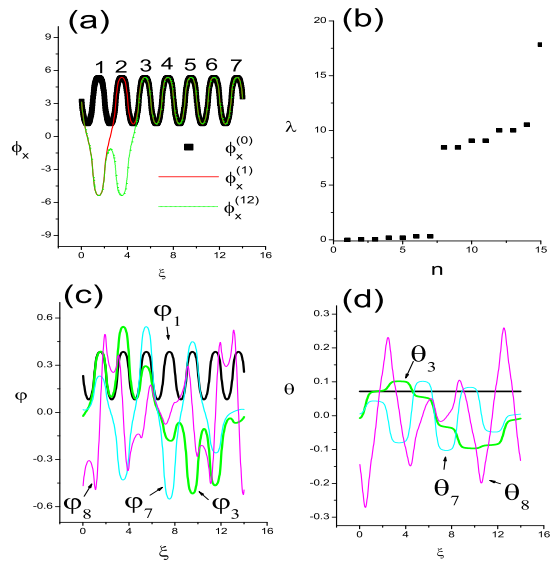}}
  \end{center}

%  \emph{
  \caption{ (Color on line)  (a) The ground state and two excited state solutions
  with temperature field with $t_w=-30,t_b=10$.
  (b) The first 15 eigenvalues of Eq. (\ref{eq:eigen}). (c) 4 typical eigen functions.
  (d) The phase variations corresponding to the eigenfunctions in (c). The straight line is $\Theta_1$.}
%  }
\end{figure}

\subsection{Gauss approximation: expansion around the ground state}

Considering the symmetry of the system, the effective Hamiltonian
should be XY type rather than Ising type. Simply let
$\sigma_i,\sigma_j$ be unit two dimensional vectors, i.e.
\begin{equation}
\sigma=(\sigma_x,\sigma_y),\hskip 1cm \sigma_x^2+\sigma_y^2=1
\label{eq:sigma}
\end{equation}
Eq. (\ref{eq:ising}) becomes a XY model. DW method does not
contradict to this conclusion and it only provided us the excited
states with $\sigma_x=\pm 1,\sigma_y=0$.

To study the cases with continuous $\sigma$, we expand the GLW
Hamiltonian near the ground state  saddle point solution. Let
\begin{equation}
\tilde{\phi}_x=\phi_x-\phi_x^{(0)}, \hskip 1cm
\tilde{\phi}_y=\phi_y-\phi_y^{(0)}=\phi_y,
\end{equation}
and we consider the Gauss approximation that
\begin{equation}
H\approx F_0+\delta H_x+\delta H_y
\end{equation}
where $F_0$ is the free energy of ground state solution and
\begin{equation}
\delta H_x=\int_0^{14}  {1 \over 2} [({d\tilde{\phi}_x \over
d\xi})^2+[t+3(\phi_x^{(0)})^2]\tilde{\phi}_x^2]d\xi
\end{equation}
\begin{equation}
\delta H_y=\int_0^{14}{1 \over 2}[({d\tilde{\phi}_y \over
d\xi})^2+[t+(\phi_x^{(0)})^2]\tilde{\phi}_y^2 ]d\xi ,
\end{equation}
where the quartic terms are omitted.

Because the saddle point solution is assumed along $\phi_x$
direction, the fluctuation of $\phi_x$ is the amplitude
fluctuation and that of $\phi_y$ is just the phase fluctuation.
The eigenmodes of $\tilde{\phi}_y$ satisfies the following
equation
\begin{equation}
-{d^2\varphi \over d\xi^2}+[t+(\phi_x^{(0)})^2]\varphi_n=\lambda_n
\varphi_n \label{eq:eigen}
\end{equation}
where $\lambda_n$ is eigenvalues and $\varphi_n$ are the
eigenfunctions. Then we have
\begin{equation}
\delta H_y={1 \over 2}\sum_n \lambda_n a_n^2 \label{eq:hy}
\end{equation}
where
\begin{equation}
a_n=\int_0^{14} \tilde{\phi}_y\varphi_n d\xi.
\end{equation}
The contribution of phase fluctuation to the partition function is
given by
\begin{equation}
z_{y}=\int D\tilde{\phi}_y e^{-\delta H_y}=\int \prod_n d a_n
e^{-{1 \over 2}\sum_n \lambda_n a_n^2} \label{eq:party}
\end{equation}

We discretize the equation (\ref{eq:eigen}) with grid of step
$0.025$ and solve it by LAPACK, which is a package to deal matrix.
The first 15 eigenvalues are shown in Fig. 1(b) and some
eigenfunctions are shown in Fig. 1(c).

The modes of $\tilde{\phi}_y$ with lower eigenvalues give the main
contribution beyond the saddle point solution. As shown in Fig.
(1b), where $t_i=-30,t_b=10$, the first 7 eigenvalues are
remarkably lower than other eigenvalues. Then they possess much
bigger thermodynamic amplitudes than other modes according to Eq.
(\ref{eq:party}).

In order to understand the first 7 eigenfunctions more clearly, we
introduce
\begin{equation}
\tan \Theta_n =\varphi_n/\phi_x^{(0)}
\end{equation}
and show $\Theta_1,\Theta_3,\Theta_7,\Theta_8$ in Fig. (1d). This
function show the phase variation of the eigenfunctions. As one
can see in Fig. (1d), $\Theta_1$ is a constant. This is because
that the first eigenfunction $\varphi_1$ has eigenvalue
$\lambda_1=0$ and satisfies
\begin{equation}
\varphi_1=\phi_x^{(0)}/\int_0^{14} d\xi (\phi_x^{(0)})^2.
\end{equation}
This can be shown by comparing the equation (3) and
(\ref{eq:eigen}). This eigenfunction corresponds to a global
rotation, so its eigenvalue is zero. $\lambda_1=0$ corresponds to
the infrared divergence.

Observing $\Theta_3,\Theta_7$, one can see that the variation of
phase in the wells are obviously smaller than in barriers.
$\Theta_2,\Theta_4,\Theta_5,\Theta_6$ also have the this feature.
From Eq.(\ref{eq:h1}), we can see that the free energy increase
related to the phase fluctuation is proportional to the square of
ground state saddle point solution. In the barriers, the saddle
point solution is much smaller than in the wells. Therefore phase
fluctuation in the barriers induced small energy increase. The
eighth eigenfuction does not have such a feature and its energy is
remarkably higher than the first 7 modes. This means that for the
first 7 modes, each block can be regarded as a unit. The phase
variation inside the block can be ignored and only the phase
difference between blocks are concerned. Therefore we introduce
the block functions
\begin{equation}
\Psi_i(\xi)=\left \{ \begin{array}{cc}
            \phi_x^{(0)}(\xi); & \phi_x^{(i)}<0, \\
            0; & other \hskip 0.5 cm cases.
           \end{array} \right.
           \label{eq:block}
\end{equation}
where the spatial range given by $\phi_x^{(i)}<0$ is explicitly
given by $2(i-1)+0.5<\xi<2i+0.5$ as mentioned in subsection III A.

Then we assume that
\begin{equation}
\tilde{\phi}_y\approx \sum_{i=1}^7 \theta_i \Psi_i,
\end{equation}
where only one phase is assigned to each block, then Eq.
(\ref{eq:hy}) becomes
\begin{equation}
\delta H_y=  \sum_{i=1}^7 \sum_{j=1}^7 J_{ij} \theta_i\theta_j
\end{equation}
where
\begin{equation}
J_{ij}={1 \over 2}\sum_{n=1}^7 \lambda_n A_{i,n}A_{j,n}
\end{equation}
with
\begin{equation}
A_{i,n}=\int_0^{14}\Psi_i(\xi)\varphi_n(\xi)d \xi.
\end{equation}
In this effective Hamiltonian, we only take the first 7 modes into
account.

On one hand the effective Hamiltonian Eq. (\ref{eq:hy1}) can be
given by the expansion of the following XY model approximately
\begin{equation}
\delta H_y \approx -\sum_{i <
j}K^{(G)}_{ij}(\cos(\theta_i-\theta_j)-1)/2
\label{eq:hy1}
\end{equation}
for $\theta_i,\theta_j \ll 1$ with
\begin{eqnarray}
J_{ii}  =  (K_{i,i-1}^{(G)}+K_{i,i+1}^{(G)})/4 ,\nonumber \\
J_{i,i+1} = J_{i+1,i}=-K_{i,i+1}^{(G)}/4, \nonumber \\
J_{i,i+2} = J_{i+2,i}=-K_{i,i+2}^{(G)}/4 \label{eq:KJ}
\end{eqnarray}
where ``(G)" is used to denote the method of Gauss approximation.
This Hamiltonian is consistent with Eq. (\ref{eq:ising}) obtained
by DW method. If we regard Eq. (\ref{eq:ising}) is a special form
of Eq. (\ref{eq:hy1}) with $\theta_i=0,\pi$, it should have
$K_{i,i+1}^{(D)}=K_{i,i+1}^{(G)}$. We investigate the cases with
different $t_b$ with fixed $t_w=-30$. The  numerical results for
matrix elements of $J_{ij}$ and the couplings $K^{(D)}_{i,i+1}$ by
DW method are shown in Table 1. As shown in table 1 and Eq.
(\ref{eq:KJ}) , $K^{(D)}_{i,i+1}\approx K_{i,i+1}^{(G)}$ is
satisfied even for $t_b=0$ in an error less than $20\%$.  For
higher $t_b$, two methods agree with each other very well.  At
$t_b=40.0$, the relative difference between $K_{i,i+1}^{(D)}$ and
$K_{i,i+1}^{(D)}$ is less than $10^{-5}$. In addition the
couplings between next nearest neighbors are much smaller than the
that between nearest neighbors, i.e. $J_{i,i+2}\ll J_{i,i+1}$.
This indicates the approximation of nearest neighbors is good
enough.

Moreover Eq. (\ref{eq:hy1}) is the well-known Josephson's
junctions Hamiltonian. This result is natural because the wells
and barriers given in Eq. (\ref{eq:lattice}) is a Josephson
junction lattice.

\begin{center}
\begin{tabular}{ccccccc}
\hline
$t_b$ & $K_{i,i+1}^{(D)}/4$ & $J_{i,i}$  & $-J_{i,i+1}$ & $J_{i,i+2}$ & $R_{\phi}$ & $R_{\lambda}$  \\
\hline
$40.0$ & $0.070354$ & $0.13905$ & $0.070357$ & $8.71\times 10^{-5}$ & $25.1$ & $549$   \\
$30.0$ & $0.16205$  & $0.32199$ & $0.16195$  & $4.70\times 10^{-4}$ & $15.5$ & $229$  \\
$20.0$ & $0.41458$  & $0.81634$ & $0.41174$  & $3.10\times 10^{-3}$  & $8.91$ & $84.5$ \\
$10.0$ & $1.2046$   & $2.2608$  & $1.15497$  & $2.50\times 10^{-2}$  & $4.72$ & $26.4$ \\
$0.0$  & $3.7826$   & $5.8779$  & $3.1027$   & $1.87\times 10^{-1}$   & $2.51$ & $7.95$ \\
\hline
\end{tabular}
\end{center}
\vskip 0.5cm {\textbf Table 1}: Couplings $K^{(D)}$ by DW method
and $J_{ij}$ by Gauss approximation at different $t_b$ and
$t_w=-30.0$.

We introduce two ratios. $R_{\phi}$ is the ratio between the
maximum of $\phi_x^{(0)}$ at the center of well and its minimum at
the center of barrier. Another ratio is defined by
$R_{\lambda}=\lambda_8/\lambda_7$. At higher $t_b$, the saddle
point solution in the barriers are very small, $R_{\phi}$ is very
large, the phase variation concentrate more in the barriers, so
the assumption is good that the phase variation inside the well is
ignored. For lower $t_b$, the saddle point solution in the
barriers is no longer small and the phase variation does not favor
concentrating in the barriers. The assumption ignoring the phase
variation inside the well is no longer good. Consider the extreme
case $t_b=t_w$, no block can be well defined.  At higher $t_b$,
the ratio $R_{\phi}$ is very large, taking only the first 7
eigenmodes and neglecting other modes is a good approximation. For
lower $t_b$, the ratio $R_{\phi}$ becomes small, the approximation
to neglect other modes becomes bad.

This approximation is similar to the phase-only approximation to
simplify the Ginzburg-Landau model to XY model \cite{bormann}, in
which the modulus of each blocks are fixed and only their phases
are allowed to fluctuate.

We summarize the Gauss approximation method as follows:

(1). Obtaining the saddle point solutions and spatial ranges of
blocks by DW method.

(2). Solving the eigenmodes of $\tilde{\phi}_y$.

(3). Defining the block functions as in Eq. (\ref{eq:block}) and
expanding the Hamiltonian  as in Eq. (\ref{eq:hy1}), then we can
get the couplings.

\subsection{Gauss approximation: expansion around the excited
state}

We can also expand the GLW Hamiltonian near the excited states
with the above method. Consider the excited state shown in Fig.
(2a). The excited state solution is obtained by assign the initial
value be negative in the 4th barrier and positive at other sites.
Therefore we denote it by $\phi_x^{(4)}$. Let
\begin{equation}
\tilde{\phi}_x=\phi_x-\phi_x^{(4)}, \hskip 1cm
\tilde{\phi}_y=\phi_y-\phi_y^{(4)}=\phi_y,
\end{equation}
and we consider the Gauss approximation that
\begin{equation}
H\approx F_4+\delta H_x+\delta H_y
\end{equation}
where $F_4$ is the free energy of excited  state $\phi_x^{(4)}$
solution and
\begin{equation}
\delta H_x=\int_0^{14} d\xi {1 \over 2} [({d\tilde{\phi}_x \over
d\xi})^2+[t+3(\phi_x^{(4)})^2]\tilde{\phi}_x^2]
\end{equation}
\begin{equation}
\delta H_y=\int_0^{14}{1 \over 2}[({d\tilde{\phi}_y \over
d\xi})^2+[t+(\phi_x^{(4)})^2]\tilde{\phi}_y^2 ],
\end{equation}
where the quartic terms are omitted.

The eigenmodes of $\tilde{\phi}_y$ satisfies the following
equation
\begin{equation}
-{d^2\varphi \over d\xi^2}+[t+(\phi_x^{(4)})^2]\varphi_n=\lambda_n
\varphi_n \label{eq:eigen1}
\end{equation}
where $\lambda_n$ is eigenvalues and $\varphi_n$ are the
eigenfunctions.

\begin{figure}
 \begin{center}
    \resizebox{8cm}{8cm}{\includegraphics{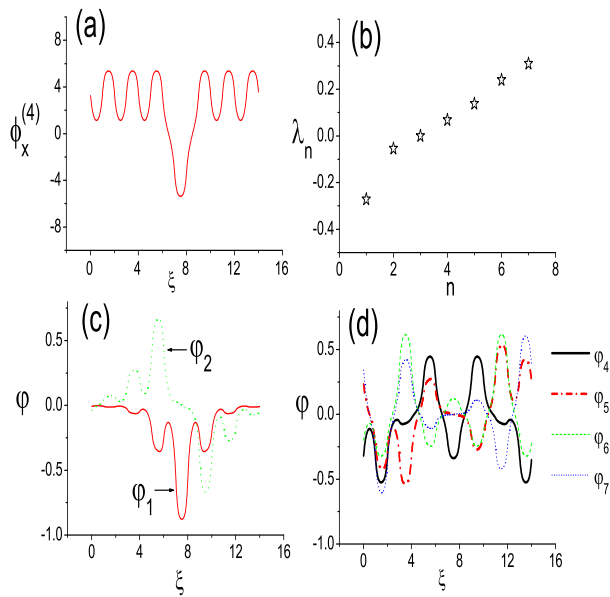}}
  \end{center}

%  \emph{
  \caption{ (Color on line)  (a) The excited state solution $\phi_x^{(4)}$. (b) The first 7 eigenvalues
  of Eq. (\ref{eq:eigen1}). (c) The eigenfunctions $\varphi_1,\varphi_2$. (d) The eigenfunctions
  $\varphi_4,\varphi_5,\varphi_6,\varphi_7$.}
%  }
\end{figure}

The first 7 eigenvalues are shown in Fig. 2(b) and some
eigenfunctions are shown in Fig. 2(c) and 2(d). The first two
eigenvalue are negative. This means that the first two modes can
cause the free energy to decrease.

The third eigenfunction $\varphi_3$ has eigenvalue $\lambda_3=0$
and it satisfies that
\begin{equation}
\varphi_3=\phi_x^{(4)}/\int_0^{14} d\xi (\phi_x^{(4)})^2.
\end{equation}
because $\phi^{(4)}$ also satisfies the saddle point equation (3)
and Eq. (\ref{eq:eigen1}) with $\lambda=0$. This mode corresponds
to a global rotation.

Similarly we introduce the block functions
\begin{equation}
\Psi_i(\xi)=\left \{ \begin{array}{cc}
            \phi_x^{(4)}; & 2(i-1)+0.5<\xi\leq 2i+0.5, \\
            0; & other \hskip 0.5cm cases.
           \end{array} \right.
\label{eq:block1}
\end{equation}
and assume that
\begin{equation}
\tilde{\phi}_y\approx \sum_{i=1}^7 \theta_i \Psi_i.
\end{equation}
Here it should be noted that the phase of the 4th block is
$\pi+\theta_4$ rather than $\theta_4$.

\begin{center}
\begin{tabular}{ccccccc}
\hline
$t_b$  & $K_{i,i+1}^{(D)}/4$ & $-J_{7,1}$ & $-J_{1,2}$  & $-J_{2,3}$ & $J_{3,4}$   \\
\hline
$40.0$ & $0.070354$          & $0.070357$ & $0.70357$ & $0.070176$  & $0.070510$    \\
$30.0$ & $0.16205$           & $0.16195$  & $0.16195$ & $0.16097$   & $0.16306$     \\
$20.0$ & $0.41458$           & $0.41174$  & $0.41174$ & $0.40510$   & $0.42221$     \\
$10.0$ & $1.2046$           & $1.1549$   & $1.1548$  & $1.0965$    & $1.2866$    \\
$0.0$  & $3.7826$            & $3.1012$   & $3.0891$  & $2.4233$    & $4.7582$     \\
\hline
\end{tabular}
\end{center}
\vskip 0.5cm {\textbf Table 2}: Couplings $K^{(D)}$ by DW method
and $J_{ij}$ by Gauss approximation at different $t_b$ and
$t_w=-30.0$ for the excited state $\phi_x^{(4)}$.

For $\theta_i \ll 1$, the effective Hamiltonian of the fluctuation
about this excited state can be expanded into following XY model
approximately
\begin{equation}
\delta H_y \approx
-\sum_{i<j}K^{(G)}_{ij}[(\cos(\theta_{0i}+\theta_i)-(\theta_{0j}+\theta_j)-1]/2
\label{eq:hy2}
\end{equation}
where
\begin{equation}
\theta_{0i}=\left \{ \begin{array}{cc}
            \pi; & i=4, \\
            0; & i\neq 4.
           \end{array} \right.
\end{equation}
Comparing Eq. (\ref{eq:hy1}) and (\ref{eq:hy2}), we get
\begin{eqnarray}
J_{3,4}=J_{4,3} = K_{3,4}^{(G)}/4, \hskip 0.5 cm J_{4,5} =J_{5,4}=
K_{4,5}^{(G)}/4, \nonumber \\
J_{i,i+1}=J_{i+1,i}=-K^{(G)}_{i,i+1}/4, \hskip 0.5 cm for \hskip
0.5cm i\neq 3,4 \hskip 0.5 cm
\end{eqnarray}

In addition to the difference between $K_{i,i+1}^{(G)}$ and
$K_{i,i+1}^{(D)}$, the lattice translational invariance is also
broken in this expansion. However for high $t_b$, the difference
between two methods becomes very small and the breaking of lattice
translational invariance also becomes very small.

We also studied the expansion near other excited states, the
conclusion is similar. Therefore we show that the effective
Hamiltonian is given by Eq. (\ref{eq:ising}) with continuous order
parameter $\sigma$ defined by Eq. (\ref{eq:sigma}).

\subsection{Gauss approximation for a real random temperature}

The periodicity in the above discussion is not essential. We apply
this method to real random temperature cases. The couplings
obtained by two methods agree with well for weak couplings.

\begin{figure}
 \begin{center}
    \resizebox{8cm}{8cm}{\includegraphics{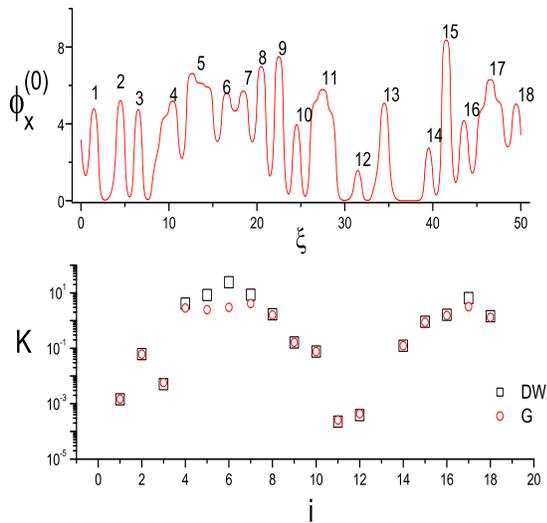}}
  \end{center}

%  \emph{
  \caption{ (Color on line)  (a) The ground state solution for
  certain random temperature realization. (b) The couplings
  between the adjoined blocks obtained by two methods. The black squares represent $K_{i,i+1}^{(D)}$
  and the red circles represent $K_{i,i+1}^{(G)}$.}
%  }
\end{figure}

We show a typical example in Fig. 2. The ground state saddle point
solution is shown in Fig. (2a). There are 18 blocks. After solving
the  Eq. (\ref{eq:eigen}) with the saddle point solution shown in
Fig. (2a), we get the eigenvalues and the eigenfunctions. We
divided the systems into 18 blocks and defined functions similar
to Eq. (\ref{eq:block}). Then we expand the Hamiltonian of the
$\phi_y$ fluctuation and get the couplings $K_{ij}^{(G)}$. The
comparison between the 18 couplings obtained by the two methods
are given Fig. (2b).

The couplings obtained by two methods agree well in the range from
$10^{-4}$ to $1$. The agreement is not good for
$K_{5,6},K_{6,7},K_{7,8}$ and $K_{16,17}$ because the saddle point
solution is not small in the regions between these couples of
blocks.

\section{Two dimensional cases}

\begin{figure}
 \begin{center}
    \resizebox{9cm}{8cm}{\includegraphics{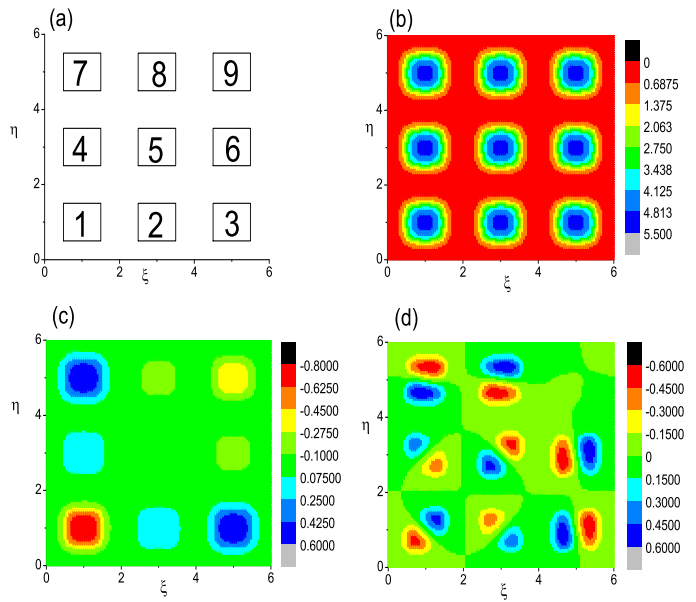}}
  \end{center}

%  \emph{
  \caption{ (Color on line)  (a) Well lattice. (b) The ground state solution.
  (c) The eigenfunction $\varphi_6$. (d) The eigenfunction
  $\varphi_{16}$. The temperatures are  $t_b=30.0,t_w=-30.0$ for (b,c,d).}
%  }
\end{figure}

We consider the well lattice  which is defined by
\begin{equation}
t(\xi,\eta)=\left \{ \begin{array}{cc}
            t_i; & \begin{array}{c} 2l-1<\xi\leq 2l \\
            2m-1<\eta\leq 2m \end{array} \\
            t_b; & other \hskip 0.5cm cases.
           \end{array} \right.
\label{eq:2d}
\end{equation}
where $l,m=1,2,3$ and $i=(m-1)*3+l$. The subscript of $t_i$ is the
label of the well, i.e., for convenience, we label the 9 wells
with $i=1,2,\cdots, 9$ as shown in Fig. 4(a).

Assume the solution along x-direction, the saddle point equation
is given by
\begin{equation}
-\frac{\partial^2 \phi_x}{\partial \xi^2}+\frac{\partial^2
\phi_x}{\partial \eta^2}+t(\xi,\eta)\phi_x+\phi_x^3=0.
\end{equation}
where the usual periodic condition is used.

Expanding the Hamiltonian about this saddle point solution, the
eigenmodes of $\tilde{\phi}_y$ satisfy
\begin{equation}
-\frac{\partial^2 \varphi_n}{\partial \xi^2}+\frac{\partial^2
\varphi_n}{\partial
\eta^2}+[t+(\phi_x^{(0)})^2]\varphi_n=\lambda_n \varphi_n
\label{eq:eigen2}
\end{equation}
where $\lambda_n$ is eigenvalues and $\varphi_n$ are the
eigenfunctions. Through the similar method given in section II, we
obtain the couplings in Gauss approximation.

We first consider the simple case with uniform  $t_i=t_w=-30.0$
for $i=1,2,3,\cdots,9$. The ground state solution for this well
lattice is shown in Fig. (4b). It can be seen that there are 9
blocks, which are the 9 cells of the well lattice. Using DW
method, we can get the couplings between these blocks. Due to the
lattice translational invariance, we can use $K_{12}$ and $K_{15}$
represent the coupling between nearest neighbors and next nearest
neighbors respectively.

\begin{center}
\begin{tabular}{ccccccc}
\hline
$t_b$ & $K_{1,2}^{(D)}/4$ & $-J_{1,2}$  & $K_{1,5}^{(D)}/2$ & $-J_{1,5}$ & $R_{\phi}$ & $R_{\lambda}$  \\
\hline
$40.0$ & $0.035649$ & $0.035720$ & $0.000234$ & $0.000199$ & $187$  & $491$   \\
$30.0$ & $0.086821$ & $0.087172$ & $0.00104$  & $0.000847$ & $96.5$ & $203$  \\
$20.0$ & $0.24358$  & $0.24549$  & $0.00591$  & $0.00455$  & $34.6$ & $72.3$ \\
$10.0$ & $0.84624$  & $0.85015$  & $0.0461$   & $0.0336 $  & $11.2$ & $19.9$ \\
$0.0$  & $3.6426$   & $3.2314$   & $0.348$    & $0.253  $  & $3.28$ & $4.75$ \\
\hline
\end{tabular}
\end{center}
\vskip 0.5cm {\textbf Table 3}: Couplings $K^{(D)}$ by DW method
and $J_{ij}$ by Gauss approximation at different $t_b$ and
$t_w=-30.0$.

As shown in Fig. (4c), for the eigenfunction $\varphi_6$, the
phase variations in the wells are much smaller than in the
barriers. The first 9 eigenfunctions have this feature and other
eigenfunctions do not have. As a example, we show $\varphi_{16}$
in Fig. (4d). And for $t_b>0$, the first 9 eigenvalues are much
smaller than other eigenvalues. Therefore we take the first 9
eigenmodes into account and obtain the couplings between the
clocks as we do in the preceding sections. The comparison between
DW method and Gauss approximation are given in Table 3 for
different $t_b$. The data in Fig. (3b, 3c, 3d) are for $t_b=30.0$.

\begin{center}
\begin{tabular}{ccccccc}
\hline

 $t_b$   &  $~~~~~~~10$ &\hskip -0.1cm $.0~~~~~~~~~$ &   $~~~~~~~20$ & \hskip -0.1cm $.0~~~~~~~~$   & $~~~~~~~~30$ & \hskip -0.1cm $.0~~~~~~~~$ \\

\hline

$ij$ & $K_{i,j}^{(D)}/4$ & $-J^{(G)}_{i,j}$  & $K_{i,j}^{(D)}/4$ & $-J^{(G)}_{i,j}$ & $K_{i,j}^{(D)}/4$ & $-J^{(G)}_{i,j}$  \\
\hline
 12 &  0.6062  &   0.6384  &   0.1610  &   0.1651  &   0.0545 &
 0.0552 \\
 23 & 0.8359  &   0.8466  &   0.2400  &   0.2429  &   0.0854 &
 0.0859 \\
 31 & 1.1650  &   1.1270  &   0.3630  &   0.3617  &   0.1358 &
 0.1359 \\
 45 & 0.6208  &   0.6649  &   0.1651  &   0.1711  &   0.0557 &
   0.0566 \\
 56 &  0.7767  &   0.8107  &   0.2189  &   0.2246  &   0.0766 &
   0.0776 \\
 64 &  1.5560  &   1.4610  &   0.5194  &   0.5114  &   0.2031 &
   0.2022 \\
 78 &  0.8220  &   0.8111  &   0.2343  &   0.2345  &   0.0831 &
   0.0832 \\
 89 &  0.3816  &   0.4314  &   0.0899  &   0.0955  &   0.0278 &
   0.0286 \\
 97 &  0.4866  &   0.5384  &   0.1217  &   0.1286  &   0.0390 &
   0.0401 \\
 14 & 1.0230  &   1.0130  &   0.3087  &   0.3099  &   0.1132 &
   0.1135 \\
 47 &  0.3634  &   0.4064  &   0.0861  &   0.0903  &   0.0269 &
   0.0264 \\
 71 &  1.7770  &   1.6400  &   0.6113  &   0.5977  &   0.2440 &
   0.2421 \\
 25 & 1.1410  &   1.1080  &   0.3522  &   0.3509  &   0.1312 &
   0.1311 \\
 58 &  0.4452  &   0.4790  &   0.1091  &   0.1136  &   0.0353 &
   0.0359 \\
 82 &  0.6657  &   0.7261  &   0.1794  &   0.1886  &   0.0604 &
   0.0619 \\
 36 & 0.9424  &   0.9379  &   0.2778  &   0.2792  &   0.1005 &
   0.1009 \\
 69 &  0.5280  &   0.5488  &   0.1358  &   0.1384  &   0.0450 &
   0.0454 \\
 93 &  0.6061  &   0.6679  &   0.1591  &   0.1679  &   0.0527 &
   0.0541 \\

\hline
\end{tabular}
\end{center}
\vskip 0.5cm {\textbf Table 4}: Couplings $K_{ij}^{(D)}$ by DW
method and $J^{(G)}_{ij}$ by Gauss approximation with nonuniform
$t_i$ at different $t_b$.

In fact only the inhomogeneity rather than the periodicity of the
temperature is essential. We also consider the random temperature
cases with $t_1,t_2,\cdots, t_9$ being
$-30,-20,-50,-40,-17,-60,-35,-25,-15$ respectively. There are $18$
couples of nearest neighbored blocks. In the table 4, we present
the couplings between nearest neighbored couples at $t_b=30.0,
20.0, 10.0$. As we can see that for $t_b=30.0$ the differences
between the couplings obtained by two methods are much smaller
than those at $t_b=10.0$.

\section{Summary}

The continuous fluctuation about the saddle point solution is
studied for XY-type Ginzburg-Landau Hamiltonian with random
temperature. The final conclusion is that

(1) The effective Hamiltonian of blocks is a XY model.

(2)The DW method provides the excited states with block's phase
being only $0,\pi$, and the Gauss approximation can describe the
continuous phase fluctuation near the saddle point solutions.

(3)The couplings obtained by these two methods agree with each
other well for the weak coupling cases.

DW method has a great advantage comparing with the Gauss
approximation. For DW method, the size of the grid in the
numerical calculation can be as large as $2000\times 2000$, while
in the Gauss approximation, the size of grid can only be as large
as $120\times 120$ and the computing time is very long because of
diagonalizing matrix. Therefore using DW method, the couplings
between blocks can be conveniently calculated and the statistical
properties of the couplings can be studied.

Our conclusion is consistent with the recent experiments and
theoretical studies. Appearance of granular structures
self-organized in homogeneously-disordered SC is discovered by the
experiment \cite{kowal} and shown by theoretical studies near the
quantum critical point  \cite{Ghosal,Dubi}.

Recently the excited state solutions of Bogliubov-de Gennes
equations are solved for two dimensional negative-U Hubbard
Hamiltonian with on-site disorder \cite{wu4}. The excited states
show that the system is self-organized into blocks. DW method is
used to obtain the couplings between blocks. The authors claimed
that the effective Hamiltonian between these blocks should be
XY-type. The argument in this paper can be regarded as its
corroborative evidence.

The author would like thank R. Ikeda for useful discussions, and
K. Noda for his help in computing. This work is supported by the
Scientific Research Foundation of State Education Ministry and the
National Basic Research Program of China (Grant No. 2007CB925004).

\end{document}